\documentclass[10pt]{iopart}
\usepackage{graphicx}

\begin{document}
\title[]{Hubbard energy dependence of electronic structures in rare-earth monoxides}
\author{Mizuki Tago, Tsukasa Kurachi, and Takayuki Makino}
\address{Graduate School of Engineering and R\&D Center for FIR Region, University of Fukui, Bunkyo 3-9-1, Fukui 910-8507, Japan}
\ead{tmakino@u-fukui.ac.jp}
\vspace{10pt}
\begin{indented}
\item[]\today
\end{indented}

\begin{abstract}
To realize the significant potential of optical materials such as strongly electron-correlated open-shell rare-earth monoxides, understanding the electron-localization Hubbard parameters ($U$) is of central importance. The Hubbard energy is believed to be material specific and constant. However, it has recently been pointed out that even subtle structural changes can induce changes in $U$ parameters. For LuO, we theoretically evaluated $U$ energy dependence of the differential transmission and reflectivity spectra to assess the impact of $U$ energy on the optical properties. In addition to the conventional derivative-like contribution, we observed the influence of the Drude tail owing to $U$-induced plasma energy modulation. Given the typical detection sensitivity of differential spectroscopy ($\sim 1 \times 10^{-4}$), even a few meV modulations in $U$ are sufficient for a detectable spectral response.

\end{abstract}

\vspace{2pc}
\noindent{\it Keywords}: strongly correlated electron system, optical properties, Hubbard parameter modulation,

\submitto{\JJAP}
%

\section{Introduction}

Lutetium monoxide (LuO) exhibits optical, electronic, and structural characteristics that are of interest in advanced technological applications and fundamental scientific studies. Concomitant with its unique dark color, LuO demonstrates high reflectance in the infrared (IR) spectral range, which may reduce IR-induced heat accumulation and holds promise for energy-efficient material applications~\cite{Kaminaga2016,Kaminaga2018a,Yamamoto2019}. Its high electrical conductivity and distinctive valence (Lu$^{2+}$: 4f$^{14}$5d$^1$) make it particularly intriguing among rare-earth compounds. Beyond these remarkable properties, the behavior of LuO as a strongly correlated electron system opens a pathway for deeper investigations into how optical excitation might influence Hubbard energy. Optical excitation is known to induce structural changes in materials via excited states. Experimental evidence for such transformations has been observed in various materials, such as alkali halides~\cite{Toyozawa1992,Toyozawa1997}, bismuth~\cite{Hase2002,Hase2005,Katsuki2013}, and oxides~\cite{SokolowskiTinten2003,Otto2018,Wall2012}, revealing the profound impact of excitation on lattice constants and crystal symmetries. These structural modifications, in turn, are closely linked to changes in Hubbard energy, a phenomenon that has been theoretically verified in systems such as FeO~\cite{Topsakal2014,Sun2020a}. However, despite the promise of such studies, the potential of optical excitation to modulate Hubbard energy via excited states has yet to be explored in rare-earth monoxides, such as lutetium monoxide (LuO).
Such modulation has the potential to drive changes in the electronic structure and dielectric functions, creating new opportunities for tailoring the optical responses in strongly correlated materials.

To delve into this unexplored frontier, this study evaluates the dependence of the electronic states and dielectric functions of LuO on the Hubbard energy ($U$) using density functional theoretical calculations. We also analyzed the differential transmission and reflectivity spectra induced by variations in Hubbard energy ($U$). This investigation aims to assess the sensitivity of $U$-induced changes in the modulation of the optical properties and dielectric responses. Given that differential spectroscopy possesses a very high sensitivity of the order of $10^{-3}$ or $10^{-4}$, even subtle modifications in $U$, which are potentially driven by external perturbations such as pressure or optical excitation, are expected to be detectable and may provide valuable insights into the underlying physical mechanisms~\cite{Versteegh2011,Nakajima2008}. This approach not only provides insights into the fundamental physics of LuO, but also opens avenues for broader applications in energy-efficient technologies and materials science.

\section{Methods}

Density functional theory (DFT)+$U$ calculations were performed using the formulation of Dudarev \textit{et al.}\cite{Dudarev1998} as implemented in the \texttt{QUANTUM ESPRESSO} code~\cite{Giannozzi2009,Giannozzi2017,Pizzi2020}. The GGA-PBE was used for
the exchange-correlation functional~\cite{Anisimov1991}. GGA is a generalized gradient approximation, whereas PBE (Perdew-Burke-Ernzerhof) is a type of hybrid functional. We used optimized norm-conserving Vanderbilt pseudopotentials~\cite{Hamann2013} for both Lu and O because we later evaluated the dielectric functions in the calculation. A kinetic energy cut-off for a charge density of 333~Ry and for wave functions of 47~Ry were used. In all cases, atomic orbitals were used to construct the occupation matrices and projectors in the PBE+$U$ scheme.

The optimized Hubbard energies were evaluated self-consistently using \texttt{hp.x}, an implementation of the density-functional perturbation theory designed to compute on-site Coulomb interaction parameters ($U$) within the DFT+$U$ framework bundled within the \texttt{Quantum ESPRESSO} suite~\cite{Floris2020,Timrov2018}. In other words, both of these values ($U_f$ and $U_p$) were adopted based on self-consistent determination, which estimates site-specific Hubbard parameters via linear response theory.
The self-consistent determination of $U$ values yielded approximately 1.5~eV for the Lu site (hereafter we call it $U_f$) and approximately 20~eV for O ($U_p$).
We found that the optimized structure has a lattice parameter of 3.878 \AA, which was adopted for the rest of the calculations.
We adopted nonzero $U_p$ values throughout this study. This is because the application of a finite Hubbard correction ($U_p$) to the O $p$states has been widely recognized as an effective approach for accurately reproducing the insulating ground state in cubic-phased CuO, thereby facilitating the opening of the band gap~\cite{Himmetoglu2011}.
For sesquioxides such as Lu$_2$O$_3$, the reference study implemented a finite Hubbard correction ($U_p$) applied to the O 2p-states, successfully reproducing a wide bandgap energy of approximately 5~eV in theoretical electronic state calculations~\cite{Jiang2012,Gillen2013,Hansen2013}. This approach further supports the validity of incorporating Hubbard energy corrections into the O 2p-states in our work.
The computational tool \texttt{epsilon.x} from the Quantum ESPRESSO suite was used to evaluate the complex dielectric function. The intra-band and inter-band smearings were set to 200 and 150 meV, respectively, to account for the broadening effects in the dielectric function calculations.

\section{Results and Discussion}

In this section, we present the results for several physical quantities, such as the density of states, complex dielectric functions, differential transmission and reflectivity spectra, and film-thickness dependence.

Figure~1 shows the density of states calculated using the PBE+$U$+$U_p$ functional. This was done to reveal the impact of varying the on-site Hubbard parameters on the electronic structure in the strongly correlated electron system.
We shifted the zero energy to the Fermi energy. As the $U$ value increases from 1.5 to 3.0~eV, we observed shifts in the optical absorption band (at ca. 5~eV) and changes in the distribution of states. A fixed $U_p$ value of 20~eV~\cite{Cococcioni2005} provides a consistent basis for comparisons across different $U$ values. 

\begin{figure}
\caption{The density of states calculated by the PBE+$U$+$U_p$ functional. The on-site Hubbard parameters are $U = $ 1.5, 2.0, 2.5, and 3.0~eV and $U_p =$ 20~eV, the latter of which is determined by the linear response approach.}
\centering
\includegraphics[width=0.7\textwidth]{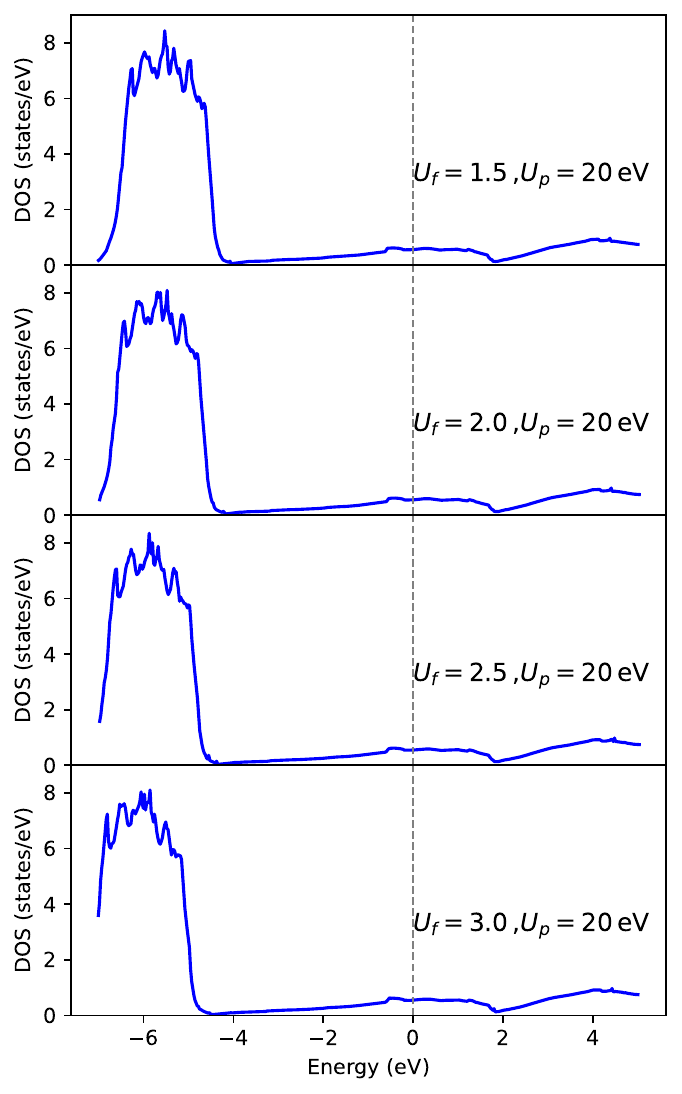}
\end{figure}

From the density of states spectra, we can conclude that the ground state of LuO is metallic irrespective of the Hubbard energy $U$ as long as we limit ourselves to this $U$ range. In a previous study~\cite{Kaminaga2018}, temperature-dependent conductivity demonstrated metallic behavior, with carrier polarity identified as the \textit{n}-type. This behavior was attributed to the 5d electrons from the Lu$^{2+}$ ions and the presence of oxygen vacancies acting as electron donors.

Figure~2 shows the dielectric functions calculated using the PBE+$U$+$U_p$ approach demonstrated the sensitivity of the optical properties to the Hubbard parameter $U$. The real and imaginary parts of the dielectric functions ($\epsilon_1$ and $\epsilon_2$) show subtle albeit detectable variations as $U$ increases from 1.5 to 3.0~eV. 

\begin{figure}
\caption{The dielectric functions calculated by the PBE+$U$+$U_p$ functional. The on-site Hubbard parameters were $U =$ 1.5, 2.0, 2.5, and 3.0~eV and $U_p =$ 20~eV.}
\centering
\includegraphics[width=0.7\textwidth]{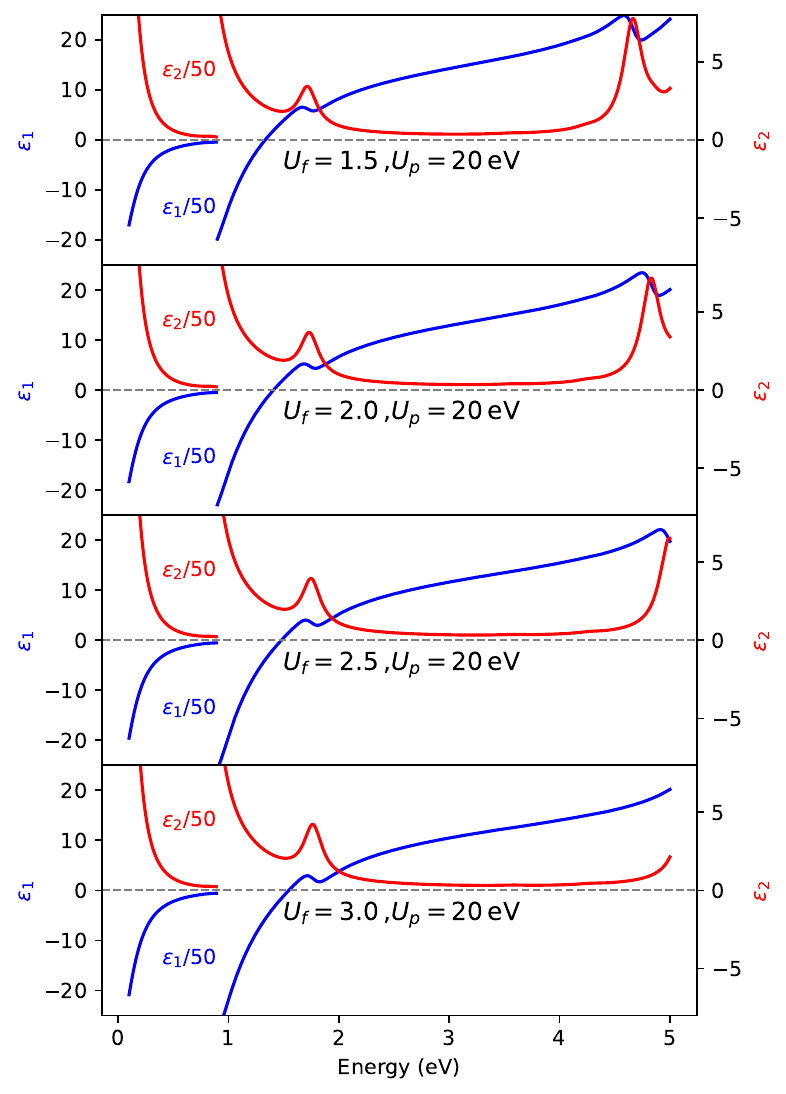}
\label{fig2}
\end{figure}

These changes in the dielectric function directly reflect $U$-induced modulation of the electronic structure and influence the optical response of the material. At the lower-energy sides, both the real and imaginary parts of the dielectric functions are characterized by Drude tails, reflecting the metallic nature of LuO's ground states. We observed a blueshift in the absorption band, which peaked at 4.8~eV for $U = 1.5$~eV, as the Hubbard $U$ energy increased. This phenomenon is attributed to the enhancement of the splitting energy induced by electron localization effects, which consequently leads to an upward shift in this band.

Because lanthanides are well-known high-k dielectric materials, a large value of $\epsilon_1$, particularly in the visible and UV energy regions of the electromagnetic spectrum, is not an outlier. At a glance, $\epsilon_2$ spectrum in the visible and UV energy regions exhibits a rather structureless line shape, but one can clearly confirm that this spectrum encompasses several distinct optical transitions (critical points) if the transmission or optical absorption spectrum is plotted. As stated earlier, we are interested in how sensibly the change in $U$ transforms to a change in transmission ($T$) or reflectance ($R$) in LuO. Before moving to the $U$-induced spectrscopic changes, we assess the sensitivity of the optical response to the Hubbard parameter $U_p$ applied to the oxygen 2p-states. To this end, transmission spectra were calculated for $U_p = 19, 20, 21 \, \mathrm{eV}$, as shown in Fig.~\ref{fighp}. The results indicate that the overall spectra remain nearly unchanged, with only a slight shift in the dip feature around the $\sim 4.8$ eV. We also confirmed that variations in $U_p$ have a much smaller impact on differential transmission and reflection compared to changes in $U_f$, supporting the robustness of the selected values.
Therefore, we calculated differential transmission and reflectance using the procedures explained below~\cite{Denton1972,Tomlin1968}. First, we evaluated the transmission and reflectivity from the complex dielectric functions. Both $T$ and $R$ are functions of complex refractive indices rather than dielectric functions. Therefore, we adopted the following formula to convert to complex refractive indices:

\begin{figure}
\caption{
Transmission spectra calculated for different oxygen Hubbard parameters $U_p \in \{19, 20, 21\} \, \mathrm{eV}$.
}
\centering
\includegraphics[width=0.6\textwidth]{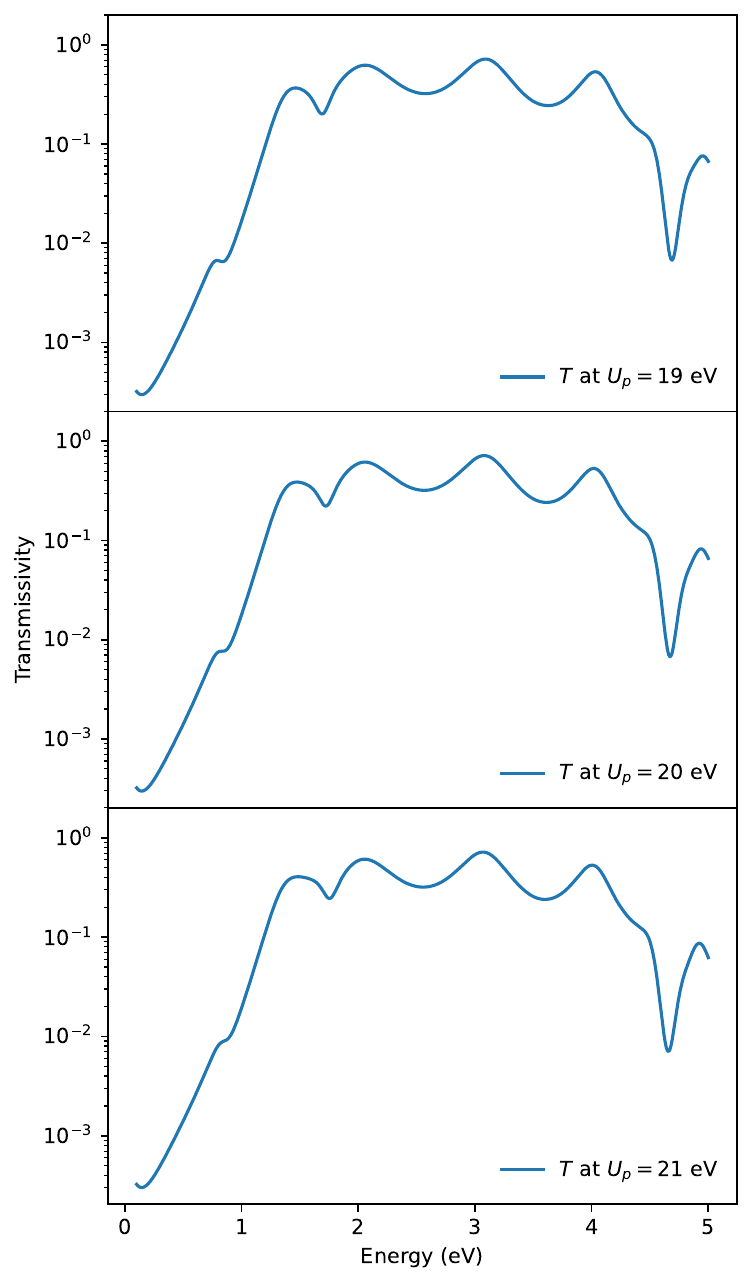}
\label{fighp}
\end{figure}

\begin{equation}
n={\frac{1}{\sqrt2}}\left(\epsilon_{1}+(\epsilon_{1}^{2}+\epsilon_{2}^{2})^{\frac{1}{2}}\right)^{\frac{1}{2}},
\end{equation}

and

\begin{equation}
\kappa ={\frac{1}{\sqrt{2}}}\left(-\epsilon_{1}+(\epsilon_{1}^{2}+\epsilon_{2}^{2})^{\frac{1}{2}}\right)^{\frac{1}{2}}.
\end{equation}

For a simple single-layer system, $R$ and $T$ are expressed as follows:

\begin{equation}
R = \frac{(n-1)^2 + \kappa^2}{(n+1)^2 + \kappa^2},
\end{equation}

and

\begin{equation}
T = (1 - R)^2 \exp{\left(- \frac{2 \kappa E \ell}{\hbar c} \right)},
\end{equation}

where $\ell$ denotes the thickness, $n$ denotes the real part of the refractive index, $\kappa$ denotes the extinction coefficient, and the others have their standard meanings.

However, pioneering work on LuO deposited this film ($\ell=105$~nm) on a CaF$_2$ substrate; therefore, we adopted the extended formula to describe the situation in a two-layer system. The explicit, very complicated equations can be found elsewhere~\cite{Denton1972}, which we used for the evaluation.

Next, to evaluate the differential spectrum, we defined the transmission intensity \( I(U,E) \) as a function of the Hubbard energy \( U \) and energy \( E \). The differential intensity was computed as follows:

\begin{equation}
\frac{\Delta I(E)}{I(E)} = \frac{I(U=2.0,E) - I(U=1.5,E)}{\frac{1}{2} \left( I(U=2.0,E) + I(U=1.5,E) \right)}.
\end{equation}

This expression quantifies the relative change in the transmission intensity between two different values of the Hubbard energy, \( U = 2.0 \) eV and \( U = 1.5 \) eV, normalized by their mean intensity. If we take the differential $I(E)$ for, for example, $U=2.5$ eV, we can replace 2.0 eV with 2.5~eV. For differential reflectivity evaluation, we adopted a procedure similar to that described above. This approach provides insight into the variation in transmission properties due to changes in the Hubbard energy and is useful for future optical experimental designs.

\begin{figure}
\centering
\includegraphics[width=0.7\textwidth]{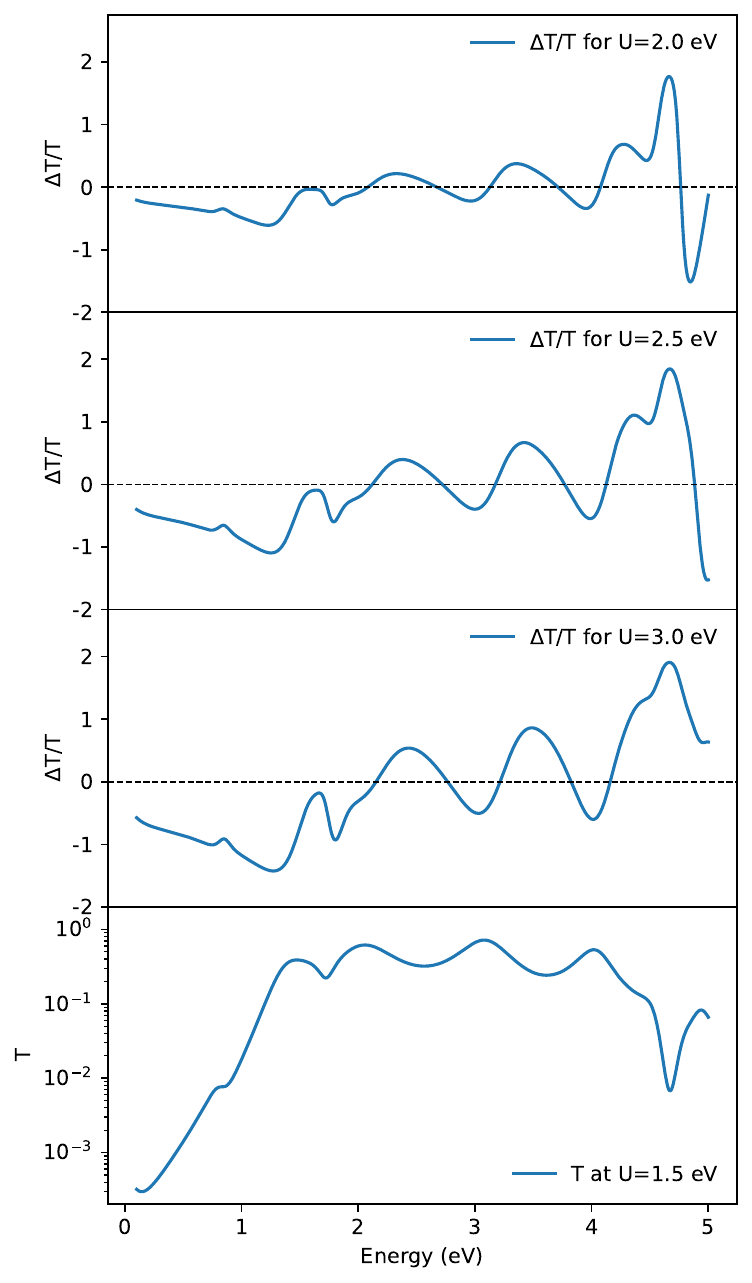}
\caption{The differential transmission spectra calculated by the PBE+$U$+$U_p$ functional. The on-site Hubbard parameters are $U =$ 1.5, 2.0, 2.5, and 3.0~eV and $U_p =$ 20~eV. The lowermost panel shows transmission ($T$) at $U=1.5$~eV and $U_p = 20$~eV in a logarithmic scale.}
\label{fig3}
\end{figure}

Figure \ref{fig3} shows the differential transmission ($\Delta T/T$) spectra computed relative to the reference $U =$ 1.5 eV case, illustrating the evolution of the optical properties with increasing Hubbard $U$ values in the PBE+$U$ framework. The transmission spectrum on a logarithmic scale at $U =$ 1.5 ~ eV is shown in the lowermost panel of Fig. \ref{fig3}.
The calculated transmission spectra revealed several distinctive peaks. A notable observation is in parallel with the reported properties of LuO thin films. The reported optical transitions are associated with the Lu 4f $\to$ Lu 5d $ t_{2g} $ transition near 1.0~eV, Lu 4f $\to$ Lu 5d $e_{g}$ transition at approximately 2.3~eV, yet an unassigned transition around 4.0~eV, and O 2p $\to$ Lu 5d $t_{2g}$ and O 2p $\to$ Lu 5d $e_{g}$ transitions near 5~eV. On the other hand, the theoretical spectrum showed transitions at \textit{ca.} 1.0~eV, 2.0~eV, 2.5~eV, 3.7~eV, 4.2~eV, 4.8~eV, respectively. If one considers that some of the splitting structures become unresolved because of the experimentally observed broadening effect, the theoretically obtained characteristic energies are in reasonable agreement with the experimental values. The calculation results are in reasonable agreement with the experimental data if we consider these calculations to be based at the DFT level. This agreement is attributed to the Hubbard correction (i.e., DFT+$U$). 

Based on these references~\cite{Kaminaga2018}, our data on dielectric functions further elucidate the electronic and optical properties intrinsic to Lu-based materials. The observed anomalies and peaks hold potential for an expanded understanding of these phenomena.
In the reference, the optical absorption spectrum lineshape was examined. Within this framework, distinguishing between the metallic and insulating ground states is challenging. Notably, the authors refrained from commenting on ground states based solely on the spectral line shape. Instead, they reinforced the assignment of metallic ground states using electrical conduction studies~\cite{Kaminaga2018}. These findings align with our calculated results, demonstrating the consistency between the observed and theoretical interpretations.

\begin{figure}
\centering
\includegraphics[width=0.7\textwidth]{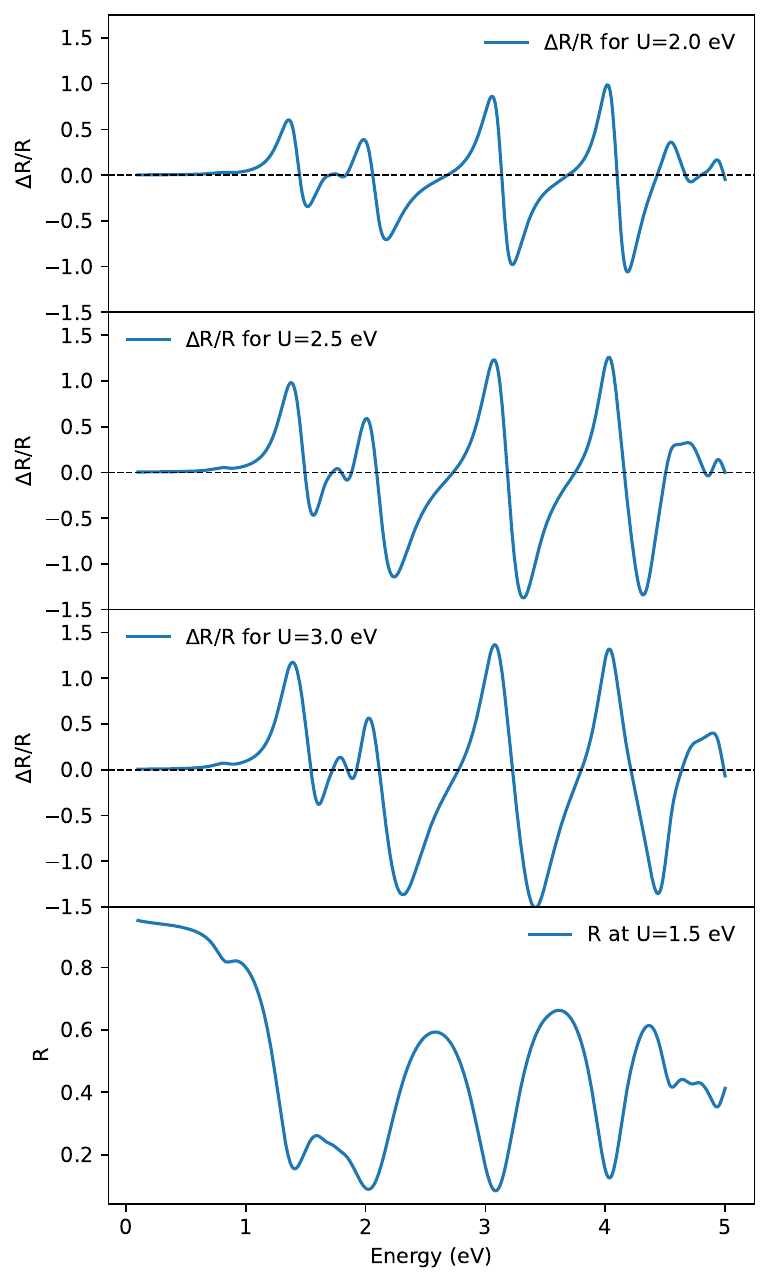}
\caption{The differential reflectivity spectra calculated by the PBE+$U$+$U_p$ functional. The on-site Hubbard parameters are $U =$ 1.5, 2.0, 2.5, and 3.0~eV and $U_p =$ 20~eV. The lowermost panel shows reflectivity ($R$) at $U=1.5$ eV and $U_p = 20$ eV.}
\label{fig4}
\end{figure}

Figure \ref{fig4} shows the differential reflectivity spectra and their dependence on Hubbard $U$ energies in the PBE+$U$+$U_p$ approach. The differential reflectivity spectra show the combined effects of varying $U$ (from 1.5 to 3.0~eV) and a fixed $U_p$ (20~eV) on optical properties~\cite{Syassen1985,Goni1990,Nayak2014}. The transmission spectrum on a logarithmic scale at $U =$ 1.5 ~ eV is shown in the lowermost panel of Fig. \ref{fig3}.
In the case of the transmission spectrum, as stated earlier, a distinctive structure appears as peaks, which are transformed into anomalies in the case of the reflectivity spectrum.

The present modulation mechanism is based on the Hubbard $U$ energy. This resulted in a change in the plasmon energies, which were deduced from the lineshape analysis of the Drude tail in the spectra. These are 5.362~eV, 5.307~eV, 5.179~eV, and 5.073~eV for $U=$ 1.5, 2.0, 2.5, and 3.0~eV. Therefore, the plasmon energies redshift with an increase in $U$ energies. On the other hand, previous studies with similar focus have investigated metallic LuH$_2$~\cite{Peterman1979,Weaver1979,Weaver1979a}. It is evident that LuO and LuH$_2$ share commonality in their open-shell electronic configurations. Prior studies have attributed the modulation mechanism to shifts in the Fermi energy~\cite{Rosei1972} and observed derivative lineshapes in the modulation spectra. The modulation mechanism in the present study, on one hand, can be attributed to the shift in plasmon energy. The lineshape in the present modulation spectra is completely different from that observed by Weaver \textit{et al.} in LuH$_2$, that is, the derivative lineshape~\cite{Peterman1979}. A detailed examination was conducted to provide deeper insight into the electron correlation effects in bad-metal systems such as LuO. A bad metal refers to a material in which electrons behave in a strongly correlated manner, and Hubbard correction effects play an important role, leading to reduced conductivity compared to normal metals.

We propose a pressure-modulated spectroscopic experiment to verify the dependence of the dielectric response on the Hubbard $U$ energy, as theoretically calculated. It is well established that variations in the lattice constants induced by external pressure lead to changes in Hubbard $U$ energy. We estimated that 1\% change in the lattice constant induces $\approx 50 \, \mathrm{meV}$ change in $U$ in LuO. Therefore, by employing a setup that applies periodic pressure to the sample, we aim to detect the resulting spectral modifications using the lock-in technique~\cite{Syassen1985}. This approach offers a typical sensitivity on the order of 10$^{-4}$. If a linear relationship between Hubbard $U$ and the differential transmittance or reflectance holds, a change in $U$ of approximately 1 meV should be sufficient for detection. For LuO, achieving such variation in $U$ requires an applied pressure of approximately 90~MPa, which is within the experimentally feasible range~\cite{Sun2020a}. Here, we supposed a bulk modulus of 150~GPa, and a pressure derivative of 4.5 (typical for rare-earth NaCl-type oxides) in the evaluation of required pressure to induce this. This supports the practicality of the proposed pressure-modulated spectroscopy experiment. If the optical excitation can change the structure via the excited state, it is highly likely to be detectable through pump-probe-type ultrafast laser spectroscopy, enabling non-invasive and non-destructive experiments.

The motivation for evaluating the thickness dependence of the differential spectroscopic value (DSV) was to determine the optimal thickness that facilitates optical experiments aimed at verifying the aforementioned simulation results. Here, DSV corresponds to $|\Delta T/T|_{max}$ and $|\Delta R/R|_{max}$. It was initially hypothesized that increasing the thickness would lead to monotonic enhancement of the DSV. However, an excessive thickness can significantly impede optical transmission, thereby introducing practical limitations. Given the interplay between the DSV enhancement and transmission constraints, a tradeoff is expected. Therefore, this study sought to identify the optimal thickness at which these competing effects are balanced, ensuring reliable spectroscopic measurements and sufficient transmission for experimental validation.

\begin{figure}
\centering
\includegraphics[width=0.75\textwidth]{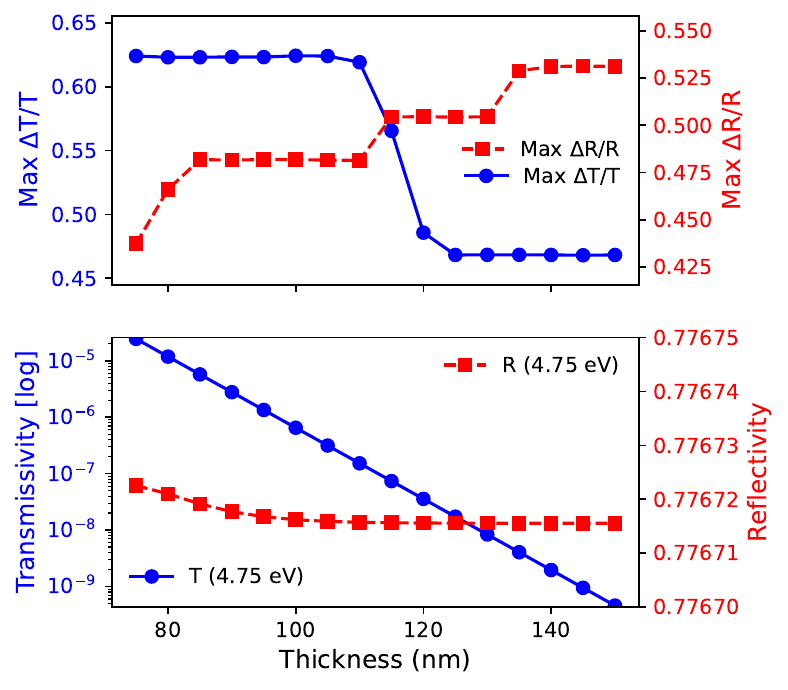}
\caption{Film thickness dependence of differential transmission and reflectivity maximum with the PBE+$U$ and the PBE+$U$+$U_p$ functionals for on-site Hubbard parameters between $U =$ 1.5 and $U_p =$ 20~eV. The bottom panel presents transmission ($T$) and reflectivity ($R$) spectra, and their thickness dependences.}
\label{fig5}
\end{figure}

Figure \ref{fig5} shows the film thickness ($\ell$) dependence of the differential transmission and reflectivity absolute maxima ($|\Delta T/T|_{max}$ and $|\Delta R/R|_{max}$) calculated using PBE+$U$+$U_p$ functionals. We also plotted transmission and reflectance spectra as a function of the film thickness. The thicknesses ranged from 50 to 150 nm. The purpose of these plots was not to provide insights into the interplay between the material thickness and Hubbard parameters ($U$), but (as stated earlier) oriented to the experimental design. The detected energy range in the spectra ranges from 0.75~eV to 4.25~eV. We chose this energy range because the transmission in the energy range below 0.75~eV is too small to detect differential transmission.
These spectra highlight the spectral regions that are most sensitive to changes in $U$, providing insight into the correlation between electronic structure modifications and optical responses. The resulting differential transmittance ($|\Delta T/T|_{max}$) exhibits a thickness dependence that appears step-like, which can be interpreted as combinations of the following multiple reasons. $|\Delta T/T|_{max}$ is composed of a single-peaked component that reflects the trade-off between enhanced excitation efficiency earned with increasing the thickness and diminished response due to reabsorption at excessive thickness, and a sinusoidally thickness-dependent component arising from optical interference. While the interference introduces a sinusoidal dependence in both transmittance and $|\Delta T/T|_{max}$, this manifests more directly in the transmittance profile. The superposition of these two effects leads to a rapid change in $|\Delta T/T|_{max}$ around a particular thickness, resulting in an overall behavior that resembles a step function.

\section{Conclusions}

This study demonstrates the effect of the Hubbard parameter $U$ on the electronic structure and optical properties of LuO. Variations in $U$ reveal a strong correlation between electron interactions and optical phenomena, including shifts in plasmon energies and absorption bands. Despite these changes, the metallic nature of the ground state of LuO persisted within the investigated $U$ range.

The computed dielectric functions and differential optical spectra highlight $U$'s role in modulating the electronic properties, reinforcing the reliability of the PBE+$U$+$U_p$ framework. By comparing theoretical predictions with prior experimental results, this study expands the understanding of $U$-induced effects in Lu-based materials. In addition, evaluating the film thickness emphasizes the need for an optimal design in spectroscopic studies.

Overall, these findings offer valuable insights into the electronic behavior of LuO and provide a foundation for future exploration of bad-metal systems and related materials.

\section*{Acknowledgments}

We acknowledge the insightful contributions of K. Kaminaga, D. Oka, and T. Fukumura, whose observations have drawn attention to the impact of localization energy on the electronic structure of this new class of materials, significantly enhancing our understanding.
The authors have benefited from the communication with I. Hamada.
Part of the computation in this study was performed using the facilities of the Supercomputer Center, the Institute for Solid State Physics, the University of Tokyo, Japan.

\textbf{Mizuki Tago}: Data curation (equal). \textbf{Tsukasa Kurachi}: Conceptualization (equal); Data curation (equal); Formal analysis (equal); Methodology (equal); Visualization (equal); Writing-original draft (supporting). \textbf{Takayuki Makino}: Conceptualization (equal); Data curation (equal); Formal analysis (equal); Methodology (equal); Investigation (lead); Supervision (lead); Validation (lead); Visualization (equal); Writing-original draft (lead); Writing-review \& editing (lead).

\section{References}
\providecommand{\newblock}{}

\end{document}